\documentclass [12pt]{article}
\usepackage{dafne99pro}
\usepackage{latexsym}
\newcommand{\beq}{\begin{equation}}
\newcommand{\eeq}{\end{equation}}
\newcommand{\beqa}{\begin{eqnarray}}
\newcommand{\eeqa}{\end{eqnarray}}

\begin{document}
\title{
HADRONIC PHYSICS
}
\author{
Ulf-G. Mei{\ss}ner\thanks{$\,\,\,$E-mail address: 
{\tt Ulf-G.Meissner@fz-juelich.de}}\\
{\em Forschungszentrum J\"ulich, Insitut f\"ur Kernphysik (Theorie)}\\
{\em D-52425 J\"ulich, Germany} \\
}
\maketitle
\baselineskip=14.5pt
\begin{abstract}
I review the novel results and developments presented at the Third
Workshop on Physics and Detectors for DA$\Phi$NE that deal with
hadronic physics. Topics discussed include: the scalar quark
condensate, kaon decays, the sector of scalar and vector mesons,
kaon-nucleon scattering, pion- and kaon-nucleon sigma terms, 
and strange nuclear physics.
\end{abstract}

\vspace{-15cm}

{\hfill {\small FZJ-IKP(TH)-00-01}}

\vspace{15cm}

\baselineskip=17pt

\section{Hadronic physics at DA$\Phi$NE energies: Why bother?}

Hadronic physics at DA$\Phi$NE covers  energies of about 1~GeV
and below. This
is a particularly challenging regime since standard perturbation
theory in the strong coupling constant $\alpha_S (Q^2)$ is not applicable. In fact,
we do not even know  from basic principles whether $\alpha_S (Q^2)$
increases monotonically with decreasing $Q^2$, as suggested by the
$\beta$--function calculated in the perturbative regime, or flattens
out. Therefore, nonperturbative methods need to be developed and employed. This is
in stark contrast to say e.g. the precise physics of the 
Standard Model tested at LEP and elsewhere. As I will discuss in
section~\ref{sec:B0}, chiral perturbation theory, eventually combined
with other methods like e.g. dispersion relations, allows one to pin
down some very fundamental parameters of QCD. These are the ratios
of the light quark masses as well as the size of the scalar
quark--antiquark condensate, which is linked to the spontaneous
symmetry violation in QCD. One can also extend these methods to
include baryons, some pertinent remarks are made in
section~\ref{sec:B1}. In particular, the so--called pion and kaon
nucleon sigma terms have attracted a lot of attention over a long
time, simply because they are the proton matrix elements of the
explicit chiral symmetry breaking part of the QCD Hamiltonian. 
In addition, this energy regime offers a rich phenomenology. For
example, it now appears that in the sector of scalar resonances, excitations have been
observed which are not simple $\bar qq$ quark model states, but have
some gluon components - either as hybrids or glueball--meson mixtures.
Many models of QCD as well as its lattice formulation (with all its
intrinsic problems) call for the existence of such states. Other 
interesting aspects of the properties of mesons in the energy range
of relevance here are also touched upon in section~\ref{sec:B0}.
Last but not least, the nucleus can act as a filter and lets us study some
processes that are forbidden in free space, one particularly
interesting example being the $\Lambda\Lambda \to \Lambda N$ 
transition which leads to the so--called non--mesonic decays of
hypernuclei. This and other recent developments are briefly surveyed
in section~\ref{sec:BN}. All the interesting new results related
to CP violation and rare kaon decays, which might hint at physics
beyond the Standard Model, are reviewed by Chris Quigg\cite{CQ}.
To summarize this brief motivation, despite many decades of studying
phenomena in the energy range accessible to DA$\Phi$NE, there are
many open questions and only recently precise theoretical tools have
been developed to answer some of these questions in a truly
quantitative manner. In addition, there is a host of new precise
data mostly related to kaon decays. Hopefully, DA$\Phi$NE will further
increase this data base soon. For other motivations and a different
point of view, I refer to Pennington's talk\cite{Penn}.

\section{The baryon number zero sector}
\label{sec:B0}
In this section, I will first make some comments on novel developments
concerning the chiral structure of QCD and then move to higher mass
states, such as the $\phi (1020)$ and the scalar sector.

\subsection{Chiral QCD}
It is well known that the QCD Lagrangian for the three light quark
flavors can be written as
\beq
{\cal L}_{\rm QCD} = {\cal L}_{\rm QCD}^0 - \bar{q} {\cal M} q~,
\eeq
where $q^T = (u,d,s)$ collects the light quark fields, ${\cal M} =
{\rm diag} (m_u,m_d,m_s)$ is the current quark mass matrix and 
the term ${\cal L}_{\rm QCD}^0$ exhibits a chiral SU(3)$_L
\times$SU(3)$_R$ symmetry. This symmetry is spontaneously broken down
to its vectorial subgroup SU(3)$_V$ with the appearance of eight
Goldstone bosons, collectively denoted as ``pions''. The pions interact
weakly at low energies. They can couple directly to the vacuum via
the axial current. The corresponding matrix element $\langle 0|A_\mu
|\pi \rangle$ is characterized by the 
typical scale of strong interactions, the pion decay constant
$F_\pi \simeq 100\,$MeV. These pions are not exactly massless but
acquire a small mass due to the explicit symmetry violation, such
as $M_\pi^2 = (m_u+m_d) B + \ldots$, where $B$
parametrizes the strength of the scalar--isoscalar quark condensate,
$B = |\langle 0|\bar{q}q|0\rangle |/F_\pi^2$. Based on these facts,
one can formulate an effective field theory (EFT) which allows one to
exactly explore the consequences of the chiral QCD
dynamics\cite{wein,gl}. This EFT is chiral perturbation theory.
Its present status has been reviewed by Gasser recently\cite{juerg}.
\subsubsection{News on the quark condensate}
Over the last few years, the question about the size of $B$
has received a lot of attention. In the standard scenario, $\langle 0|
\bar{q}q|0\rangle \simeq (-230\,{\rm MeV})^3$, so that $B \simeq
1.4\,$GeV and one can make very precise predictions, as reviewed here
by Colangelo\cite{col}. In particular, the isospin zero S--wave
$\pi\pi$ scattering length $a_0^0$  can be predicted to better than 5\%
accuracy. However, the value of $B$ might be smaller. In fact, one
can reorder the chiral expansion allowing to float $B$ from values
as small as $F_\pi \simeq 100\,$MeV to the standard case\cite{jan}. For a small value
of $B$, the quark mass term has to be counted differently and to a
given order in the chiral expansion, one has more parameters to pin
down. For $B$ on the small side, $a_0^0$ could be as much as 30\%
larger than in the standard case. These two scenarios lead also to a significant
difference in the quark mass expansion of the Goldstone bosons. Consider e.g.
the charged pions,
\beq
M_{\pi^\pm}^2 = (m_u + m_d) B + (m_u + m_d)^2 A + {\cal O}(m_{u,d}^3)~.
\eeq
In the standard scenario the linear term is much bigger than the
quadratic one, in the large $B$ case they are of comparable size. An
immediate consequence is that while in the first case the Gell-Mann--Okubo
relation $4M_K^2 = 3M_\eta^2 + M_\pi^2 $ comes out naturally, in the other
scenario parameter tuning is necessary.  For a discussion of what can be
learned from lattice gauge theory in this context, see e.g. the lectures 
by Ecker\cite{Eck}. Ultimately, this question has to be
decided experimentally. So far, the best ``direct'' information on the
S--wave $\pi\pi$ scattering phase close to threshold comes from $K_{\ell 4}$
decays, since due to the final--state theorem of Fermi and Watson, the
phase of the produced pion pair is nothing but $\delta_0^0 (s) -
\delta_1^1 (s)$ with $\sqrt{s} \in [280,380]$~MeV and $\delta_1^1 (s)
< 1^\circ$ in this energy range. All data from the seventies seem to
indicate a large scattering length with an sizeable error. This
unsatisfactory situation will be improved very soon.
The preliminary data from the BNL E865 collaboration were
shown by J. Lowe\cite{JL} (for a glimpse on these data, see the contribution
of S. Pislak to HadAtom 99\cite{HA99}). They are not yet final, in particular
radiative corrections have not yet been accounted for, but taken face
value, they are clearly supporting the standard scenario. 
\subsubsection{Pionic atoms}
Another
method to measure the elusive S--wave scattering length comes from
the lifetime of $\pi^+\pi^-$ atoms. This electromagnetic bound state
with a size of approximately 400~fm can interact strongly and decay
into a pair of neutral pions. The lifetime of this atom is directly proportional
to the S--wave scattering length difference $|a_0^0 -
a_0^2|^2$. Therefore, a determination of this lifetime to 10\% gives
the scattering length difference to 5\%. The DIRAC experiment at the
CERN SPS is well underway as reported by Adeva\cite{Dirac}. Also,
the theory is well under control. Recent work by the Bern
group\cite{bern} has lead to a very precise formula relating the
lifetime to $\pi\pi$ scattering including isospin breaking in the
light quark mass difference and the electric charge (the formalism is
developed in refs.\cite{MMS,KU}). It is mandatory that the experimenters
use this improved Deser--type formula in their analysis!
It would also be interesting to
calculate the properties of $\pi K$ atoms and measure their lifetime.
For a much more detailed discussion I refer to the proceedings of
HadAtom~99\cite{HA99}.
\subsubsection{Kaon decays}
As stressed in the talks by D'Ambrosio\cite{DAm} and
Colangelo\cite{col}, there are many chiral perturbation
theory predictions for all possible kaon decay modes. It was
therefore very interesting to see that a huge amount of new data
is available and still to come, as detailed in the talks of Lowe\cite{JL},
Kettell\cite{ket} and Flyagin\cite{fly}. For the sake of brevity,
I will only discuss three topics here. 
\begin{itemize}
\item \underline{$K^0_L \to \pi^0 \gamma\gamma$:} 
This is a particularly
interesting decay with a long history. It vanishes at leading order 
${\cal O}(p^2)$ in
the chiral expansion and is given by a finite loop effect at
next-to-leading order, ${\cal O}(p^4)$.
While the predicted two--photon spectrum~\cite{EPR} agreed
well with the data~\cite{Kdat}, the branching ratio was underestimated by
about a factor of three. To cure that, unitarity corrections and
higher order contact terms have been considered. In particular, at
order $p^6$ there is an important vector--meson--dominance
contribution, parametrized in terms of the coupling $a_V$. The ${\cal
  O}(p^6)$ calculation with $a_V = -0.7$ not only improves the
two--photon spectrum but also the branching ratio agrees with
experiment. More important, as stressed by D'Ambrosio, this value for
$a_V$ is consistent with a VMD model and analysis of the process
$K_L \to \gamma\gamma^*$.
\item \underline{$K \to \pi \gamma^*$:} 
This decay mode was discussed by
d'Ambrosio and Lowe. The matrix element for this process is given in
terms of one invariant function, $A(K\to\pi l^+l^-) \sim W(z)$, with
$z = (M_{ll}/M_K)^2$ and $M_{ll}$ the mass of the lepton pair.
The invariant function $W(z)$ has the generic form
\beq\label{W}
W(z) = \alpha + \beta z + W_{\pi\pi} (z)~,
\eeq
where $\alpha$ and $\beta$ are related to some low--energy constants,
but the momentum dependence of the pion loop contribution $W_{\pi\pi} (z)$
is unique
and leads to unambiguous prediction. The
data shown by Lowe can indeed be described significantly better
with the form given in eq.(\ref{W}) than with a linear polynom with
also two free parameters. Thus, we have another clear indication of
chiral pion loops. 
\item \underline{$K \to 3\pi$:} 
The non-leptonic weak
chiral Lagrangian has a host of undetermined parameters at
next-to-leading order. For specific reactions, like e.g. $K \to 2\pi$
or $K\to 3\pi$, only a few of these enter. It is thus important to have
some data to pin down these constants and based on that, make further
predictions. Flyagin showed some results from SERPUKHOV on the mode
$K^+ \to \pi^+ \pi^0 \pi^0$. In terms of slope and quadratic slope
parameters, the invariant matrix element squared can be written as
$|M|^2 \sim 1+gX+hX^2 + kY^2$, with $X,Y$ properly scaled relative
pion momenta. The three slopes $g,h$ and $k$ could be determined and
thus further tests of the weak non-leptonic chiral Lagrangian are possible. 
\end{itemize}

\subsection{Higher masses}

In the region between 1 and 2 GeV, the spectrum of states is
particularly rich and interesting. As explained in detail by  Barnes\cite{TB}
and  Donnachie\cite{AD}, we now have some first solid evidence for {\it
  glueballs} and {\it hybrids}. Glueballs are states made of glue with
no quark content. In a ideal world of very many colors, $N_C \to
\infty$, the glueball sector
decouples from the sector made of mesons and baryons, i.e. the states
made of quarks and anti--quarks, see refs.\cite{tHooft}\cite{Witten}. In
the real world with $N_C =3$, matters are more complicated. The decay
pattern of the glueball candidate as mapped out in big detail by the
Crystal Barrel collaboration\cite{CB} is most simply interpreted in
terms of mixing, most probably of two genuine meson and one glueball state.
Similarly, there are evidences for hybrids, i.e. states made of quarks
and ``constituent'' gluons, a particularly solid candidate being the
$1^{-+} (\rho\pi) (1600)$.\footnote{Notice that it is important that
such states have ``exotic'' quantum numbers. If not, one can always
cook up some minor modifications of the quark model to explain states
with constituent gluons by some other mechanism. One quite old example
is debated in refs.\cite{CH,UGM}.} Clearly, if one such state exists, there
is no reason to believe that there are not many more (Pandora's box?). In particular, 
DA$\Phi$NE could contribute significantly to the search for vector
hybrids like the $\phi ' \sim |s\bar s g\rangle$ or the $\omega '$ 
-- if these are not too heavy. After these more general remarks, 
let me turn to two special topics.

\subsection{Remarks on the scalar sector}
The scalar meson sector is still most controversial. It consists of
the elusive ``sigma'', the $a_0$, the $f_0$ and so on. Much debate
is focusing about the nature of these states, which of them belong
to the quark model octet/nonet (assignment problem), which of these are
$K \bar K$ molecules (structure problem) and so on. Certainly, these
scalars can be produced in photon--photon fusion at DA$\Phi$NE.
I will not dwell on these issues here but rather add some opinion about 
the the ``sigma'', which is labeled $f_0 (400-1200)$ by PDG. First, 
a ``charming'' new result was reported by  Appel\cite{Appel} 
in one parallel session. The invariant mass distribution of the final
state of the decay $D^+ \to \pi^+
\pi^0 \pi^0$ measured at FNAL was analyzed in terms of conventional
resonances and could not be explained. If one adds, however, a $\pi
\sigma$ contribution, this turns out to be a strong channel and the $\sigma$
parameters from a best fit are $M_\sigma = 486\,$MeV and  
$\Gamma_\sigma = 351\,$MeV, in agreement with other interpretations
of $\pi \pi$ scattering data, for a recent review see e.g.\cite{Pennsig}.
The role of such a state in the $\phi \to \pi^+ \pi^- \gamma$ decay was
discussed here by  Lucio\cite{JLuc}. I would like to take the
opportunity to add my opinion about this state:
\begin{itemize}
\item It is not a ``pre--existing'' resonance, but  rather a dynamic
effect due to the strong pion--pion interaction  in the isospin zero,
S--wave. Specific examples how to generate such a light and broad
sigma are the modified Omn\`es resummation in chiral perturbation
theory~\cite{GM,UGMcomm} or the chiral unitary approach of Oller and
Oset\cite{OO,Osettalk}, or others.
\item It is certainly {\it not} the chiral partner of the pion, as
suggested by models based on a linear representation of chiral
symmetry. For a critical analysis of the renormalizable
$\sigma$--model in the context of QCD, I refer to ref.\cite{gl}.
\item It is long known in nuclear physics that the intermediate
range attraction between two nucleons can be explained by the exchange
of a light sigma. It is also known since long how to generate such a
state in terms of pion rescattering and box graphs including
intermediate delta isobars, for a nice exposition see e.g. ref.\cite{KH}.
\end{itemize}
I was particularly amazed to see the many new and interesting data
from $e^+ e^-$ annihilation at VEPP--2M (Novosibirsk), which were presented by
Salnikov\cite{AS}  and Milstein\cite{AM1}. I will only pick out
three aspects of these results, which I found most interesting:
\begin{itemize}
\item The three pion final state $\pi^+ \pi^- \pi^0$ indicates the
existence of a low--lying $\omega'$ mesons at $M_{\omega '} = (1170
\pm 10)\,$MeV with a width of $\Gamma_{\omega '} = (197 \pm 15)\,$MeV.
Also confirmed is the $\omega ' (1600)$, whereas the $\omega ' (1420)$
was not seen. The role of low--lying (effective) excited omegas in the
analysis of the strange vector currents and the violation of the OZI  
rule is discussed e.g. in ref.\cite{MMSO}.  
\item The analysis of the decays $\phi \to f_0 \gamma, a_0 \gamma,
\eta \pi \gamma$ lends credit to the hypothesis that the $a_0$ and
$f_0$ are $qq\bar q \bar q$ and not simple $q\bar q$ states.
\item The channel $e^+ e^- \to 4\pi$ is dominated by the $a_1 (1260) \pi$
intermediate state. The $a_1 \pi$ amplitude extracted by the
Novosibirsk group from electron--positron annihilation\cite{AM2} 
is completely consistent with the one obtained from analyzing the high
precision data on $\tau \to 3\pi \nu_\tau$ from CLEO and ALEPH\cite{taudata}. 
\end{itemize}

\section{The baryon number one sector}
\label{sec:B1}

I now turn to processes involving exactly one baryon in the initial
and the final state. Of most relevance for DA$\Phi$NE is, of course,
the kaon--nucleon system. However, before one can hope to tackle this
problem in a truly quantitative manner, it is mandatory of having 
obtained a deep understanding of the somewhat ``cleaner'' pion--nucleon
system. This refers to a) the smallness of the up and down quark
masses compared to the strange quark mass, which makes explicit
symmetry breaking easier to handle (i.e. a faster convergence of the
chiral expansion) and b) to the appearance of very close to or even
subthreshold resonances in the KN system, like e.g. the famous 
$\Lambda (1405)$ -- such interesting complications do not arise 
in pion--nucleon scattering. Before considering explicit examples,
we should address the following question:
 
\subsection{What can we learn?}
Clearly, the chiral structure of QCD in the sector with baryon number
one is interesting {\it per} {\it se}. Some prominent examples which
have attracted lots of attention are neutral pion photoproduction, real and 
virtual Compton scattering off the proton or hyperon radii and
polarizabilities, to name a few. In all these cases, the relevance of
chiral pion loops is by now firmly established and underlines the
importance of the pion cloud for the structure of the ground state
baryons in the non--perturbative regime. The analysis of the baryon
mass spectrum allows to give further constraints on the ratios of the
light quark masses, see e.g. ref.\cite{juergM,bora}. Furthermore,
in the pion--nucleon system, isospin breaking $\sim (m_u -m_d)$ and
explicit chiral symmetry  $\sim (m_u +m_d)$ start at the same order,
quite in contrast to the pion case. In addition, much interest has
been focused on the question of ``strangeness in the nucleon'', more
precisely the expectation values of operators containing strange
quarks in nucleon states. The sigma terms discussed below are
sensitive to the scalar operator $\bar s s$. Complementary information
can be obtained from parity--violating electron scattering ($\sim \bar s
\gamma_\mu s$) or polarized deep inelastic lepton scattering ($\sim \bar s
\gamma_\mu \gamma_5 s$).

\subsection{Lessons from $\pi$N}
It is important to recall some lessons learned from pion--nucleon
scattering (in some cases the hard way). As emphasized in the clear 
talks by Gasser\cite{juergT} and Rusetsky\cite{AR}, not only is
the scalar sector of chiral QCD intrinsically difficult but also for
making precise predictions at low energies, one has to consider
strong and electromagnetic isospin violation besides the hadronic
isospin--conserving chiral corrections. Often, it is mandatory to
combine chiral perturbation theory with dispersion relations to
achieve the required accuracy. As a shining example, I recall the
pion--nucleon sigma term story (a very basic and clear introduction
using the {\it pion} sigma term as a guideline is given in Gasser's
talk\cite{juergT}). The quantity that one wants to determine is
\beq
\sigma (t=0) = \langle p | \hat m (\bar u u + \bar d d)|p\rangle~,
\eeq
with $|p\rangle$ a proton state of momentum $p$, $\hat m$ is the
average light quark mass and $t$ the invariant momentum transfer squared. 
Clearly, momentum transfer zero is not
accessible in the physical region of $\pi$N scattering. So how can one
get to this quantity? The starting point is the venerable
low--energy theorem of Brown, Pardee and Peccei\cite{BPP}
\beq
\Sigma  = \sigma (0) + \Delta\sigma + \Delta_R~.
\eeq
Here, $\Sigma = F_\pi^2 \bar{D}^+ (\nu =0,
t=2M_\pi^2)$ is the
isoscalar $\pi$N scattering amplitude with the pseudovector Born
term subtracted at the Cheng--Dashen point\footnote{This point in the
Mandelstam plane is special because chiral (pion mass) corrections 
are minimal.}, and $M_\pi$ and $F_\pi$ are the charged pion mass and 
the weak pion decay constant, respectively. 
The numerical value of $\Sigma$ can be obtained by
using hyperbolic dispersion relations and the existing pion--nucleon
scattering data base. The most recent determination of $\Sigma$ based
on this method is due to Stahov\cite{JS}, $\Sigma = 65 \ldots
75\,$MeV, not very different from the much older Karlsruhe analysis.
The scalar form factor, $\Delta \sigma = \sigma (2M_\pi^2)
- \sigma (0)$ has been most systematically analyzed in ref.\cite{GLS}.
The resulting value of $\Delta\sigma \simeq 15\,$MeV translates into
a huge scalar nucleon radius of $r_S^2 \simeq 1.6\,$fm$^2$ (note that
the typical electromagnetic nucleon radii are of the order of
0.7~fm$^2$). 
A similar enhancement of the scalar radius also appears for the pion, see e.g.
refs.\cite{gl,GM}. Finally, $\Delta_R$ is a remainder not fixed by 
chiral symmetry. The most systematic evaluation of this quantity
has lead to an upper bound, $\Delta_R \simeq 2\,$MeV\cite{BKMcd}.
Putting all these small pieces together, one arrives at $\sigma (0)
\simeq (45 \pm 10)\,$MeV which translates into $y = 2\langle p | \bar
ss |p \rangle / \langle p | \bar u u + \bar d d| p \rangle \simeq 0.2
\pm 0.1$. These results have been confirmed recently using a quite
different approach\cite{Paul} (using also the Karlsruhe--Helsinki
phase shift analysis as input). This determination of $\Sigma$ has
been challenged over the
years by the VPI/GW group (and others). Their most recent number is
sizeably larger, $\Sigma \simeq 90 \pm8\,$MeV\cite{Marcello}. However,
if one employs the method of ref.\cite{Paul} to the $\bar{A}^+$ amplitude of
the latest two VPI/GW partial analyses (SP99 and SM99), one gets
a much larger sigma term, $\sigma (0) \simeq 200\,$MeV. This casts
some doubts on the internal consistency of the VPI/GW analysis. 
Personally, I do  not understand how such a large value for the
sigma term could be made consistent with other implications of chiral
dynamics in the meson--baryon sector. In this context, I also wish to
point out that so far, we have considered an isospin symmetric world.
In ref.\cite{sven} it was shown that isospin violation can amount to 
a 8\% reduction of $\sigma (0)$  and Rusetsky\cite{AR} demonstrated that the
electromagnetic corrections used so far in the analysis of pionic
hydrogen to determine the S--wave scattering length\cite{PSI} have
presumably been underestimated substantially. The moral is that to
make a precise statement in this context, many small pieces have to
be calculated precisely. Committing a sin at any place leads to a
result which should not be trusted.
Finally, I mention that astrophysical consequences of the  strange scalar
 nucleon  matrix element are discussed in ref.\cite{WIMP}. 

\subsection{Status and perspectives for KN}
After this detour, I come back to kaons, i.e. the kaon--nucleon system
as discussed by  Olin\cite{OL}, touched upon by Gasser\cite{juergT}
and for a recent review, see ref.\cite{gensini}. Because of the
strange quark, one can form two new sigma terms, which are labelled 
$\sigma_{KN}^{(1,2)}$ in the isospin basis or $\sigma_{KN}^{(u,d)}$ in
the quark basis,
\beqa
\sigma^{(1)}_{KN} (t) &=& \frac{1}{2} (\hat m + m_s) \langle p' | \bar uu + 
\bar ss | p \rangle~, \nonumber \\ 
\sigma^{(2)}_{KN} &=& \frac{1}{2} (\hat m + m_s) \langle p' | -\bar uu + 
2\bar dd + \bar ss  | p \rangle~,
\eeqa
with $t=(p'-p)^2$. These novel sigma terms in principle encode the
same information about $y$ as does the pion--nucleon sigma term. This
is one reason for attempting to determine them. One also needs to know the
kaon--nucleon scattering amplitude as input for strangeness nuclear
physics, as discussed in the next section. So there is ample need to
improve the data basis and obtain a better theoretical understanding. 
I briefly review where we stand with respect to low--energy
kaon--nucleon interactions.

\subsubsection{Status report}
I begin with a  summary of the data, as reviewed by Olin\cite{OL}. 
Consider first $K^+ N$.
For total isospin $I=1$ (obtained from elastic $K^+ p$ scattering),
the S--waves are fairly well known and the P--waves are small. The
situation for the $I=0$ data based on $K^+ d$ scattering and $K_L^0 p
\to K^+ n$ is very unsatisfactory - the S--waves are very uncertain
and the P--waves are very large already at small momentum. This is the
equivalent channel to the isoscalar S--wave $\pi$N amplitude, i.e.
to leading chiral order (current algebra) the pertinent scattering
length vanishes. $K^- N$ is, of course, resonance dominated due to the
presence of the strange quark. The most famous state here is the $\Lambda
(1405)$, which has been interpreted by some as a KN subtreshold
(virtual) bound state whereas others consider it a ``normal'' three
quark state. Clearly, such very different pictures should lead to
very pronounced differences in the electromagnetic radii or other
observables. These two pictures can eventually be
disentangled by electroproduction experiments. How that can work
has been shown for the $S_{11} (1535)$ in ref.\cite{FJSSW}, where
it was demonstrated that electroproduction off deuterium, $e+d \to
e' + N + N^*$, can be sensitive to the structure of the resonance 
$N^*$ under consideration. Data on $K^0 N$ are not very precise.
There is also information on the $K^- p$ bound state. The long
standing discrepancy between the data from kaonic hydrogen and
extrapolation of $KN$ scattering data to zero energy was resolved
by the fine experiment at KEK\cite{KEK}. The strong interaction shift
turned out to be negative and also the width could be determined,
but not very precisely. 
\subsubsection{Prospects for DA$\Phi$NE}
The DEAR experiment, which was discussed by
Guaraldo\cite{Carlo}, attempts to determine the strong interaction
shift and width of kaonic hydrogen to an accuracy of 1\% and 3\%, 
respectively. If that will be achieved, it 
would essentially pin down zero energy S--wave scattering and become
a benchmark point. Beware, however, that to determine the KN sigma
terms much more precise information (coming from scattering) will be
needed. Also, the theoretical analysis needs to be sharpened since the
KN Cheng--Dashen point at $t=4M_K^2 \simeq 1\,$GeV$^2$ is very far away from the
zero energy point. As stressed by Olin\cite{OL}, FINUDA will attempt
to measure $K_L^0 p$ scattering reactions to 5\% accuracy, however,
in a fairly small momentum interval. The good news is that the
theoretical machinery has considerably improved over the last years.
First, the rigorous work by the Bern group on $\pi^+ \pi^-$ and $\pi^-
p$ bound states\cite{bern,AR} can certainly be extended to the $K^-
p$ case (for that, a detailed investigation of electromagnetic
corrections for $K\pi$ scattering has to be done -- and is underway\cite{dilb}). 
Second, KN scattering has been considered based on SU(3) chiral
Lagrangian using coupled channel techniques\cite{Munich,Valencia}. 
In these approaches, one uses chiral symmetry to constrain the
potentials between the various channels and with a few parameters
(some from the chiral Lagrangian and other from the regularization),
one can describe a wealth of data related to scattering, decays and
also electromagnetic reactions. It would still be interesting to
implement even stronger constraints on the KN system, such as the leading
Goldstone boson loop effects. One particularly interesting outcome
of these studies is that not only the $\Lambda (1405)$ but also the
$S_{11}(1535)$ are quasi--bound $\bar K N$ and $K^+ Y$ states,
respectively (as mentioned above). So it appears that more precise
data as expected from DA$\Phi$NE are timely and will contribute
significantly to our understanding of three flavor meson--baryon dynamics.

\section{The baryon number greater than one sector}
\label{sec:BN}
I now turn to the nucleus, more precisely, to systems with more than
one nucleon. The objects to be studied are hypernuclei, i.e. nuclei
with one (or more) bound hyperon(s) (or even cascades) and also
atomic and nuclear kaonic bound states. This is the realm of
what is often called strangeness nuclear physics\footnote{I prefer
to call it {\it strange nuclear physics} because of  the many ``strange'',
that is: interesting, phenomena happening in such systems.}. Before discussing
some specific examples, we have to address the following question:

\subsection{Why ``strange'' nuclear physics?}
The properties of hypernuclei are of course sensitive to the
fundamental $YN$ and $YY$ (for strangeness $S=-2$) interactions.
A solid determination of interactions in such systems  allows one e.g.
to address the question of flavor SU(3) symmetry in hadronic
interactions. Furthermore, one can study the weak interactions
of baryons in the nuclear medium. Of special interest are novel
mechanism like $\Lambda N \to NN$, which have $\Delta S =1$ and
have parity conserving as well as parity violating components. This
might eventually give some novel insight into the $\Delta I =1/2$ rule.
Electromagnetic production of hypernuclei is complementary to the
usual hadronic mechanisms like e.g. stopping of kaons and thus
one can access different levels and get a more complete picture
of hypernuclear properties. One can also study the $\bar K N$
effective interaction or the kaon--nucleus interaction at rest in
deeply bound kaonic states. Mesons and baryons with strangeness
can also affect the nuclear equation of state significantly and
thus might lead to interesting phenomena in astrophysics and
relativistic heavy ion collisions. For these reasons (and others),
an intense experimental program is underway or upcoming at 
KEK\cite{Nag,Ima}, BNL\cite{Ima}, Dubna\cite{Luk},
TJNAF\cite{Iod} and DA$\Phi$NE\cite{Zen}, COSY and other labs.

\subsection{Example~1: Non-mesonic decays of hypernuclei}
Spectroscopy of $\Lambda$--hypernuclei allows one to study the
fundamental $\Lambda N$ interaction. The weak decays of such nuclei
give additional tests of elementary particle physics theories, as
discussed in the talk by Ramos\cite{angels}. In free space, the $\Lambda$
decays into $p \pi^-$ and $n \pi^0$, with a relative branching fraction 
of about 2. This is another manifestation of the $\Delta I = 1/2$ rule.
In typical nuclei, the Fermi momentum is about 300~MeV, i.e. larger than
nucleon momentum in the free $\Lambda$ decay, $p_N \simeq 100\,$MeV. 
Thus, the mesonic decay is
Pauli blocked and new decay channels open, like  the one--nucleon induced 
decay, $\Lambda n \to nn$ and $\Lambda p \to p$ with the corresponding
partial width $\Gamma_n$ and $\Gamma_p$, respectively.  Another non--mesonic
channel is the 2N--induced decay, $\Lambda np \to nnp$.
In the one--pion-exchange (OPE) model, one can
describe roughly the total non--mesonic decay rate, but for that one has
to include form factors at the vertices as well as to account for the
strong $\Lambda N$ and $NN$ interactions in the final and initial state,
respectively. The form factor dependence is particularly troublesome,
since in a truly field theoretic description of one--boson--exchange,
such a  concept makes no sense. Also, in OPE tensor transitions are enhanced,
which lets one expect that  $\Gamma_n/\Gamma_p $ is small, quite in contrast
to the experimental finding  $\Gamma_n/\Gamma_p \simeq 1$. As shown by Ramos,
the inclusion of other mechanisms like exchanges of heavier mesons,
correlated two--pion exchange or the two-nucleon induced decay
do not resolve this problem. Even worse, calculations within seemingly equivalent
models lead to very different results for the partial rates. So it seems
mandatory to develop better models, based e.g. on the latest Nijmegen 
$YN$ potential or the upcoming improved J\"ulich model\cite{johan}.
I would like to issue two warnings here: First, as already remarked,
the area of meson--exchange models supplemented by form factors is
certainly at its end, more systematic effective field theory approaches
will eventually take over. Such a change of dogma is presently happening
on the level of the $NN$ force. Second, it should also be stressed that
very few is known about the underlying $YNM$ couplings - this has been
stressed in another context in ref.\cite{jan2}.

\subsection{Example~2: $\Lambda\Sigma^0$ mixing effects}
An important effect in $\Lambda$--hypernuclei is the mixing of
the $\Lambda$ with the $\Sigma^0$. Consequences of this mixing were
discussed by Akaishi\cite{Aka} and Motoba\cite{Mot}.
It solves e.g. the overbinding
problem in $^5_\Lambda$He, which was pointed out by Dalitz and 
others\cite{dal} long time ago. The $0^+$ level in  $^5_\Lambda$He
moves to the correct binding energy due to the transition potential
$V_{\Lambda N, \Sigma N} (Q/e) V_{\Sigma N, \Lambda N}$ taken e.g.
from the Nijmegen potential (version D). Here, the operator $Q$ assures
the Pauli principle and the energy denominator $e$ deviates from its free
space version $e_0$ due to energy dissipation. It was also pointed out
by Motoba that the  $\Lambda\Sigma^0$ coupling in the  $0^+$ states of
$^4_\Lambda$H and $^4_\Lambda$He is significantly enhanced due to coherent
addition of various components, which leads to a very strong and 
attractive $NNN\to NN\Lambda$
three--body force. Of course, all these findings are very sensitive to the
underlying $YN$ interaction, which can not yet be pinned down  very reliably
due to the lack of sufficiently many precise data. 

\subsection{Other interesting results}
There were many other interesting developments, I just mention three
examples:
\begin{itemize}
\item Friedman\cite{Eli} described work on deeply bound kaonic atomic
states, which can be calculated by use of an optical potential,  
$V_{\rm opt}$. It was
demonstrated that if this optical potential is obtained from a fit to the
existing kaonic atom data, the predictions for the deeply bound states
are independent of the precise form of $V_{\rm opt}$. These states can best
be produced by the $(\phi,K^+)$ reaction for $p_\phi \simeq 170$~MeV (which
can e.g. be achieved in an asymmetric $e^+e^-$ collider).
\item Motoba\cite{Mot} and Imai\cite{Ima} discussed the possible role
of the $\Lambda$ as ``glue'' in the nucleus, leading to a shrinkage of
nuclear radii. A particular example is  $^7_\Lambda$Li, which in a cluster
model can be described by an alpha--particle plus $\Lambda$--``core'' surrounded
by a neutron--proton pair. From the measurement of E2 and M1 transitions,
one can deduce the radius, which indeed turns out smaller than the one
of the equivalent system composed of nucleons only.
\item As discussed by Imai\cite{Ima}, the H--dibaryon simply does not
want to show up. Even after a long term dedicated effort to find this
six quark state, no signal has been found. Despite its uniqueness, it
seems to have the same fate as all predicted dibaryon -- nonexistence.
\end{itemize}

\section{Expectations for the next DA$\Phi$NE workshop}

With KLOE, FINUDA and DEAR hopefully soon producing data with the
expected precision and experiments at other laboratories also
supplying precision data, we can expect to discuss significant progress
in our understanding of hadronic physics in the GeV region. On the
theoretical side, apart from all the surprises to come, I mention a
few topics which need to and will be addressed (this list is meant in no
way to be exhaustive but rather reflects some of my personal preferences):
\begin{itemize}
\item In two as well as three flavor meson chiral perturbation theory,
hadronic two loop calculations have been performed for a variety of
processes. It has, however, become clear that at that accuracy one also
needs to consider electromagnetic corrections. For the kaon decays to
be measured at DA$\Phi$NE and elsewhere, such calculation must also include
the leptons. The corresponding machinery to perform such investigations
is found in ref.\cite{emlep}. 
\item The calculation of the properties of hadronic atoms has received
considerable attention over the last years, triggered mostly by the
precise data from PSI for pionic hydrogen and deuterium and the DIRAC
experiment (``pionium''). The effective field theory methods, which have proven so
valuable for these systems, should be extended to the cases of $\pi^- K^+$ 
and $\pi^- d$ bound states to learn more about SU(3) chiral symmetry and the
isoscalar S--wave pion--nucleon scattering length, respectively.
\item Better models, eventually guided by lattice gauge theory, are needed
to understand the structure of the observed exotic states and scalar mesons.
It would be valuable to combine the quark model with constraints from
chiral symmetry and also channel couplings. Only then a unique interpretation
of these states can be achieved. Needless to say that besides the spectrum
one also has to calculate decay widths and so on.
\item A new dispersion--theoretical analysis of the pion--nucleon scattering
data, including also isospin breaking effects (beyond the pion, nucleon and
delta mass splittings) is called for to get better constraints on the pion--nucleon
scattering amplitude in the unphysical region and thus pin down the sigma term
more reliably. Presently available partial wave analyses are not including
sufficiently many theoretical constraints (or are based on an outdated data set).
\item Chiral Lagrangian approaches to low energy kaon--nucleon interactions
should be refined. So far, the necessary resummation methods start from
the leading or next--to--leading order effective Lagrangian. Thus, only
certain classes of loop graphs are included. I consider it mandatory to
also include the leading effects of the meson cloud consistently. How this
can be done in the (much simpler) pion--nucleon system is demonstrated in
ref.\cite{MO}.
\item The fundamental hyperon--nucleon interaction, which is not 
only interesting {\it per}
{\it se} but also a necessary ingredient for the calculation of hypernuclei,
has to be studied in more detail. As already mentioned, the J\"ulich group
is presently working on a refined meson--exchange model\cite{johan}. I also
expect studies based on effective field theory to give deeper insight, for
a first step see ref.\cite{sasp}.
\end{itemize}

\section*{Acknowledgements}
I would like to thank the organizers, in particular Giorgio Capon, 
Gino Isidori and Giulia Pancheri, for setting up such an interesting
meeting and making me attend all talks. Warm thanks also to 
the secretaries and staff, especially Laura Sirugo, for all their
efforts and help. 


%

\begin{thebibliography}{99}
\bibitem{CQ} C. Quigg, ``CP violation and rare decays'', {\it these proceedings.}
\bibitem{Penn} M. Pennington, ``Low energy hadronic physics'', {\it these proceedings.}
\bibitem{wein} S. Weinberg,  Physica  96A (1979) 327.
\bibitem{gl} J. Gasser and H. Leutwyler, Ann. Phys. (NY) 158 (1984) 142.
\bibitem{juerg}J. Gasser, ``Chiral Perturbation Theory'', {\tt hep-ph/9912548}. 
\bibitem{col}G. Colangelo, ``Chiral perturbation theory: an
  overview'',  {\it these proceedings.}
\bibitem{jan}see e.g. J. Stern, H. Sazdjian and N. Fuchs,
Phys. Rev.  D47 (1993) 3814.
\bibitem{Eck}G. Ecker, ``Chiral symmetry'', {\tt hep-ph/9905500}.
\bibitem{JL}J. Lowe, ``Experimental results on semileptonic K decays and
form factors'', {\it these proceedings.}
\bibitem{HA99} Mini-Proceedings of HadAtom~99,
(eds. J. Gasser, A. Rusetsky and J. Schacher, Bern, 1999),
{\tt hep-ph/9911332}.
\bibitem{Dirac}B. Adeva, ``The Dirac experiment at CERN'', {\it these proceedings.}
\bibitem{bern} A. Gall, J. Gasser, V.E. Lyubovitskij and A. Rusetsky,
Phys. Lett. B462 (1999) 335.
\bibitem{MMS} Ulf-G. Mei{\ss}ner, G. M\"uller and S. Steininger,
Phys. Lett. B406 (1997) 154; (E) Phys. Lett. B407 (1997) 454.
\bibitem{KU} M. Knecht and R. Urech, Nucl. Phys. B519 (1998) 329.
\bibitem{DAm}G. D'Ambrosio, ``Radiative rare K decays'', {\it these proceedings.}
\bibitem{ket}S. Kettell, ``Experimental results on radiative and 
non--leptonic K decays'', {\it these proceedings.}
\bibitem{fly}V. Flyagin, ``Latest results on K decays from
  Serpukhov'', {\it these proceedings.}
\bibitem{EPR}G. Ecker, A. Pich and E. de Rafael,
Nucl. Phys.  B303 (1988) 665.
\bibitem{Kdat}G.D. Barr et al.,  Phys. Lett. B242 (1990) 523;
V. Papadimitrou et al., Phys. Rev.  D44 (1991) 573.
\bibitem{TB}T. Barnes, ``Spectroscopy of light mesons'', {\it these proceedings.}
\bibitem{AD}A. Donnachie, ``Hybrid mesonic states, their relevance
to DA$\Phi$NE'', {\it these proceedings.}
\bibitem{tHooft} G. 't Hooft, Nucl. Phys. B72 (1974) 461.
\bibitem{Witten} E. Witten, Nucl. Phys. B160 (1979) 57.
\bibitem{CB} The Crystal Barrel Collaboration (V.V. Anisovich et al.),
Phys. Lett. B323 (1994) 322, and many following papers.
\bibitem{CH} C.E. Carlson and T.H. Hansson, Phys. Lett. 128B (1983) 95.
\bibitem{UGM} Ulf-G. Mei{\ss}ner, Phys. Lett. 128B (1983) 99.
\bibitem{Appel}J.A. Appel, ``Light mesons through charm decays'',
{\it these proceedings.}
\bibitem{Pennsig}M. Pennington, ``Riddle of the scalars: where is the
  sigma?'', {\tt hep-ph/9905241}. 
\bibitem{JLuc}J.L.  Lucio, ``Effects of the $f_0 (400-1200)$ in $\phi \to
\pi^+ \pi^- \gamma$'', {\it these proceedings.}
\bibitem{GM}J. Gasser and Ulf-G. Mei{\ss}ner, Nucl. Phys. B357 (1991)
  90.
\bibitem{UGMcomm} Ulf-G. Mei{\ss}ner, Comm. Nucl. Part. Phys. 20 (1991) 119. 
\bibitem{OO}J.A. Oller and E. Oset, Phys. Rev. D60 (1999) 074023.
\bibitem{Osettalk} E. Oset, ``Radiative $\phi$ decays and $\phi \to
\pi^+ \pi^- $ in a chiral unitary approach'', {\it these proceedings.}
\bibitem{KH}K. Holinde, Prog. Part. Nucl. Phys. 36 (1996) 311, and references therein.
\bibitem{AS} A. Salnikov,``Review of experimental results from  SND
detector at VEPP-2M'', {\it these proceedings.}
\bibitem{AM1} A.I. Milstein,``Review of experimental results from CMD-2
detector at VEPP-2M'', {\it these proceedings.}
\bibitem{MMSO} Ulf-G. Mei{\ss}ner, V. Mull, J. Speth and J.W. Van
  Orden, Phys. Lett. B408 (1997) 381. 
\bibitem{AM2} A.I. Milstein,``$a_1 \pi$ contribution to $e^+ e^- \to
  4\pi$ at VEPP-2M'', {\it these proceedings.}
\bibitem{taudata}H. Albrecht et al., Phys. Lett. B260 (1991) 259; 
D. Busculic et al., Zeit. f. Physik C74 (197) 263;
R. Balaest et al., Phys. Rev. Lett. 75 (1995) 3809;
M. Athanas et al., ``Limit on $\tau$ neutrino mass from $\tau \to  \pi^-
\pi^+ \pi^0 \nu_\tau$, {\tt hep-ex/9906015}.
\bibitem{juergM}J. Gasser, Ann. Phys. (NY) 136 (1982) 62.
\bibitem{bora} B. Borasoy and Ulf-G. Mei{\ss}ner, Ann. Phys. (NY) 254 (1997) 192.
\bibitem{juergT}J. Gasser, ``Sigma term physics'',  {\it these proceedings.}
\bibitem{AR}A. Rusetsky, ``Hadronic atoms in QCD'',  {\it these proceedings.}
\bibitem{BPP} L.S. Brown, W.J. Pardee and R.D.  Peccei,
  Phys. Rev. D4 (1971) 2801.
\bibitem{JS} J. Stahov, talk given at MENU'99 (Zuoz, Switzerland,
  August 1999).
\bibitem{GLS} J. Gasser, H. Leutwyler and M.E. Sainio, 
  Phys. Lett. B253 (1991) 252.
\bibitem{BKMcd}V. Bernard, N. Kaiser and Ulf-G. Mei{\ss}ner,
  Phys. Lett. B389 (1996) 144.
\bibitem{Paul}P. B\"uttiker and Ulf-G. Mei{\ss}ner. ``Pion--nucleon
scattering inside the Mandelstam triangle'', {\tt hep-ph/9908247},
to appear in Nucl. Phys. A.
\bibitem{Marcello} M.M. Pavan, R.A. Arndt, I.I Strakovsky and
  R.L. Workman, ``New Result for the Pion-Nucleon Sigma Term from an 
 Updated VPI/GW Pion-Nucleon Partial-Wave and  Dispersion Relation
 Analysis'', {\tt nucl-th/9912034}.
\bibitem{sven}Ulf-G. Mei{\ss}ner and S. Steininger, Phys. Lett. B419 (1998) 403.
\bibitem{PSI} H.-Ch. Schr\"oder et al., Phys. Lett. B469 (1999) 25.
\bibitem{WIMP}A. Bottino, F. Donate, N. Fornengo and S. Scopel,
``Implications for relic neutralinos of the theoretical uncertainties
in the neutralino--nucleon cross section'', {\tt hep-ph/9909228}.
\bibitem{OL}A. Olin, ``K-N scattering and K-N low energy
  interactions'', {\it these proceedings.}
\bibitem{gensini}P. Gensini, ``KN sigma terms, strangeness in the
  nucleon and DA$\Phi$NE'', {\tt hep-ph/9804344}. 
\bibitem{FJSSW} L. Frankfurt et al, Phys. Rev. C60 (1999) 055202.
\bibitem{KEK}M. Iwasaki et al., Phys. Rev. Lett. 78 (199è) 3067.
\bibitem{Carlo}C. Guaraldo, ``DEAR physics program'', {\it these proceedings.}
\bibitem{dilb} B. Kubis and Ulf-G. Mei{\ss}ner, forthcomimg.
\bibitem{Munich} N. Kaiser, P.B. Siegel and W. Weise, 
Nucl. Phys. A594 (1995) 325.
\bibitem{Valencia} E. Oset and A. Ramos, Nucl. Phys. A635 (1998) 99. 
\bibitem{Nag}T. Nagae, ``Recent results and perspectives from the
SKS spectrometer'', {\it these proceedings.}
\bibitem{Ima}K. Imai, ``Hypernuclear physics at BNL and KEK'',
{\it these proceedings.}
\bibitem{Luk}J. Lukstins, ``Hypernuclear physics program at Dubna'',
{\it these proceedings.}
\bibitem{Iod}M. Iodice, ``Hypernuclear physics at Jefferson Lab'',
{\it these proceedings.}
\bibitem{Zen}A. Zenoni, ``FINUDA physics program'',
{\it these proceedings.}
\bibitem{angels}A. Ramos, ``The non-mesonic weak decay of 
$\Lambda$--hypernuclei'', {\it these proceedings.}
\bibitem{johan}J.  Haidenbauer, W. Melnitschouk and J. Speth,
 {\tt nucl-th/9805014}. 
\bibitem{jan2}J. Stern, H. Sazdjian and N. Fuchs, Phys. Lett. B238 (1990) 380.
\bibitem{Aka}Y. Akaishi, ``Strangeness in nuclear matter'', 
{\it these proceedings.}
\bibitem{Mot}T. Motoba, ``Status and perspectives of hypernuclear physics'',
{\it these proceedings.}
\bibitem{dal}R. Dalitz et al., Nucl. Phys. B47 (1972) 109.
\bibitem{Eli}E. Friedman, ``Searching for deeply bound $K^-$ atomic states
at DA$\Phi$NE'',  {\it these proceedings.}
\bibitem{emlep}M. Knecht, H. Neufeld, H. Rupertsberger and P. Talavera, 
 ``Chiral perturbation theory with virtual photons and leptons'',
{\tt hep-ph/9909284}.
\bibitem{MO}Ulf-G. Mei{\ss}ner and J.A. Oller, {\tt nucl-th/9912026}.
\bibitem{sasp} M.J. Savage and R.P. Springer, Nucl. Phys. A639 (1998) 325.
\end{thebibliography}
\end{document}